\documentclass[12pt]{article}
\usepackage{amsmath} 
\usepackage{pstricks}
\usepackage{longtable}
\usepackage[dvips]{graphicx}       
\usepackage{color}
\usepackage{cite}
\input{epsf} 
\newcommand\as{\alpha_{\mathrm{S}}} 
\newcommand\f[2]{\frac{#1}{#2}}

\def\beq{\begin{equation}} 
\def\eeq{\end{equation}} 
\def\to{\rightarrow} 
\def\nn{\nonumber}

\def\b0{\beta_0}

\def\beeq{\begin{eqnarray}}
\def\eeeq{\end{eqnarray}}
\def\ep{\epsilon}
\def\bom#1{{\mbox{\boldmath $#1$}}}

\def\mur{\mu_R} 
\def\muf{\mu_F}
\def\mur2{\mu_R^2} 
\def\muf2{\mu_F^2}

\begin{document}

\begin{titlepage}
\renewcommand{\thefootnote}{\fnsymbol{footnote}}
\begin{flushright}
\end{flushright}
\par \vspace{10mm}

\begin{center}
{\Large \bf

Two-loop virtual corrections to Higgs pair production
}
\end{center}
\par \vspace{2mm}
\begin{center}
{\bf Daniel de Florian}\footnote{deflo@df.uba.ar} and
{\bf Javier Mazzitelli}\footnote{jmazzi@df.uba.ar}\\

\vspace{5mm}

Departamento de F\'\i sica, FCEyN, Universidad de Buenos Aires, \\
(1428) Pabell\'on 1, Ciudad Universitaria, Capital Federal, Argentina\\

\vspace{5mm}

\end{center}

\par \vspace{2mm}
\begin{center} {\large \bf Abstract} \end{center}
\begin{quote}
\pretolerance 10000

We present the two-loop virtual corrections to Standard Model Higgs boson pair production via gluon fusion $gg\to HH$ in the heavy top quark limit. 
Based on this result, we evaluate the corresponding
cross section at the LHC at $14$ TeV in the next-to-next-to-leading order soft-virtual approximation. 

We find an inclusive {\it K}-factor of about $2.4$, resulting in an increase close to $23\% $ with respect to the previous available calculation at next-to-leading order. As expected, we observe a considerable reduction in the renormalization and factorization scale dependence.

\end{quote}

\vspace*{\fill}
\begin{flushleft}
\end{flushleft}
\end{titlepage}

\setcounter{footnote}{1}
\renewcommand{\thefootnote}{\fnsymbol{footnote}}

\section{Introduction}

Recently, both ATLAS and CMS collaborations have discovered a new boson with a mass around $125\,\text{GeV}$ \cite{Aad:2012tfa,Chatrchyan:2012ufa} at the Large Hadron Collider (LHC). Its properties are, so far, compatible with the long sought Standard Model (SM) Higgs boson \cite{Englert:1964et}.
In order to decide whether this particle is indeed responsible for the Electroweak Symmetry Breaking (EWSB), it is crucial to measure its couplings to fermions and gauge bosons and to verify their proportionality to the particle masses. Furthermore, a precise measurement of the Higgs self-interaction is needed.

The measurement of the Higgs self-couplings is the only way to reconstruct the scalar potential. 
After EWSB, the Higgs potential takes the form
\beq
V(H)=\f{1}{2}M_H^2 H^2+\lambda\, v H^3+\f{1}{4}\lambda' H^4\,.
\eeq
In the SM the trilinear and quartic self-couplings take the same value, $\lambda=\lambda'=M_H^2/(2v^2)$, where  $v\simeq 246\,\text{GeV}$ is the Higgs vacuum expectation value and $M_H$ its mass.
In most new physics scenarios these couplings deviate from the SM values. Therefore, a  determination of the Higgs self-interaction is necessary both to understand the EWSB mechanism and to try to distinguish the SM from other models.

The Higgs quartic coupling can be in principle studied via triple Higgs boson production. However, this cross section is too small to be measured at the LHC \cite{Plehn:2005nk}, and then a determination of its value is not possible at present time. The situation is different for the trilinear coupling $\lambda$ via Higgs pair production if very high luminosities can be achieved,

The possibilities of observing Higgs pair production at the LHC have been discussed in Refs. \cite{Baur:2002qd,
Dolan:2012rv,Papaefstathiou:2012qe,
Baglio:2012np,Baur:2003gp,Dolan:2012ac,Goertz:2013kp,
Shao:2013bz}.
Though the analysis is challenging due to the smallness of the signal cross section and the large QCD background, it has been shown to be achievable at a 
luminosity-upgraded LHC.
For example for $b\bar{b}\gamma\gamma$ and $b\bar{b}\tau^+\tau^-$ final states, after the application of proper cuts, the significances obtained are $\sim 16$ and $\sim 9$ respectively, for $\sqrt{s_H}=14\,\text{TeV}$ and $\int {\cal L}=3000\,\text{fb}^{-1}$ \cite{Baglio:2012np}. These are so far the most promising final states for the Higgs trilinear coupling analysis.
The application of jet substructure techniques was shown to be important to further improve on the sensitivity of the discovery channels \cite{Dolan:2012rv,Papaefstathiou:2012qe,Gouzevitch:2013qca}.

As it occurs for single Higgs \cite{Georgi:1977gs}, the dominant mechanism for SM Higgs pair production at hadron colliders is gluon-gluon fusion, mediated by a heavy-quark (mainly top) loop. The corresponding cross section has been calculated at leading-order (LO) in Refs. \cite{Glover:1987nx,Eboli:1987dy,Plehn:1996wb}. The next-to-leading order (NLO) QCD corrections have been evaluated in Ref. \cite{Dawson:1998py} in the large top-mass approximation and found to be rather large, with an inclusive {\it K}-factor close to $2$, a very similar situation to the one observed for single-Higgs production at the same order \cite{Dawson:1990zj,Djouadi:1991tk,Spira:1995rr}. Considering that the  
next-to-next-to-leading order (NNLO) corrections for single-Higgs are also sizable \cite{Harlander:2002wh,Anastasiou:2002yz,Ravindran:2003um}, it becomes essential to reach the same accuracy for double-Higgs production in order to provide precise predictions for the process.

A full NNLO calculation requires the evaluation of the corresponding amplitudes for double real radiation, real emission from one-loop corrections and the pure virtual two-loop contribution.
In this article we present the explicit results for two-loop virtual corrections to the partonic process $gg\to HH$ in the heavy top quark limit. Furthermore, we combine these results with the universal formula presented in Ref. \cite{deFlorian:2012za} to obtain the NNLO soft-virtual approximation to the cross section, as a first step towards a full NNLO calculation,
and present numerical results for the cross section expected at the LHC within that approximation.

\section{Two-loop virtual corrections}

We present here our results on the two-loop corrections. In order to simplify the presentation we directly provide the contribution of two-loop diagrams to the corresponding partonic cross section. As usual, divergences are dealt with by using dimensional regularization with $n=4-2\ep$ dimensions, and we use the $\overline{\text{MS}}$ renormalization scheme.

As it was mentioned before, we strictly  work within the heavy top quark approximation, where the single and double-Higgs coupling to gluons is given by the effective Lagrangian
\beq
{\cal L}_{\text{eff}}=
-\f{1}{4}G_{\mu\nu}G^{\mu\nu}\left(C_H\f{H}{v}-C_{HH}\f{H^2}{v^2}\right)\,,
\eeq
where $G_{\mu\nu}$ represents the gluonic field strength tensor. In order to obtain the NNLO cross section for $gg\to HH$, we need the coefficients $C_H$ and $C_{HH}$ up to ${\cal O}(\as^3)$. The first one takes the following form \cite{Kramer:1996iq,Chetyrkin:1997iv}:
\beq\label{CH}
C_H=-\f{1}{3}\f{\as}{\pi}\left\{
1+\f{11}{4}\f{\as}{\pi}+
\left(\f{\as}{\pi}\right)^2\left[
\f{2777}{288}+\f{19}{16}\log\f{\mu_R^2}{M_t^2}+N_f\left(
-\f{67}{96}+\f{1}{3}\log\f{\mu_R^2}{M_t^2}
\right)
\right]
+{\cal O}(\as^3)
\right\}\,,
\eeq
where $M_t$ is the on-shell top quark mass, $\mu_R$ is the renormalization scale and $N_f$ is the number of light flavors. The coefficient $C_{HH}$ is known up to ${\cal O}(\as^2)$ \cite{Djouadi:1991tk}, and coincides to that order to $C_H$. We will write
\beq
C_{HH}=-\f{1}{3}\f{\as}{\pi}\left\{
1+\f{11}{4}\f{\as}{\pi}+
\left(\f{\as}{\pi}\right)^2
C_{HH}^{(2)}
+{\cal O}(\as^3)
\right\}\,.
\eeq
For the phenomenological results, we will assume $C_{HH}^{(2)}=C_{H}^{(2)}$, where the latter is defined by the squared bracket in Eq.(\ref{CH}).

\begin{figure}
\vspace{-0.5cm}
\begin{center}
\begin{tabular}{c}
\epsfxsize=8truecm
\epsffile{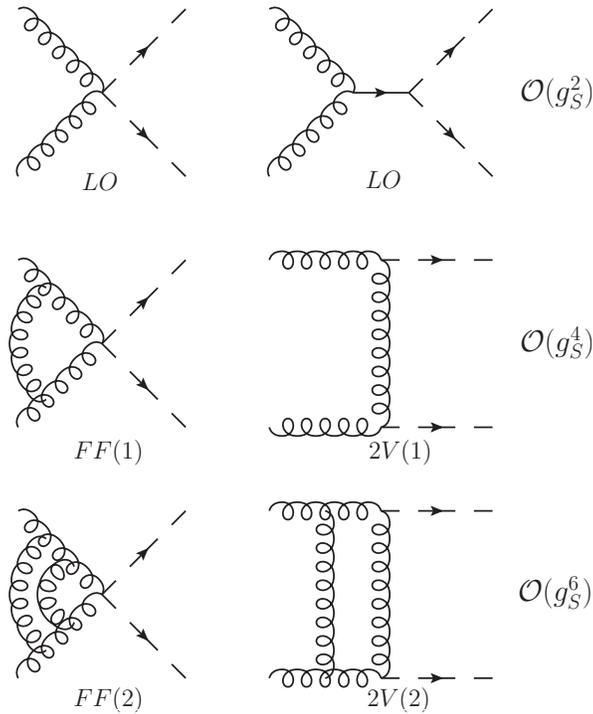}\\
\end{tabular}
\end{center}
\vspace{-0.5cm}
\caption{\label{diagramas}
A sample of the Feynman diagrams needed for the double-Higgs NNLO virtual corrections, and the corresponding label for each kind of contribution.}
\end{figure}

In Figure \ref{diagramas} we show a sample of the Feynman diagrams needed for the calculation, and we introduce the notation for each contribution. Since the structure of $gHH$ and $ggHH$ vertices is the same, the loop corrections to both Born level diagrams are proportional to the gluon form factor, and those contributions are labeled as FF(1) and FF(2). To compute these amplitudes, we rely on the two-loop gluon form factors presented in \cite{Harlander:2000mg,Gehrmann:2005pd,Baikov:2009bg,Gehrmann:2010ue}.
On the other hand, the contributions arising from diagrams with tree-level two $gHH$ vertices (labeled as 2V(1)) and the corresponding one-loop correction to them (labeled as 2V(2)), which are of the same order in powers of the strong coupling constant as the form factor-like corrections FF(1) and FF(2) have more complex kinematics and require an explicit computation.

The NNLO virtual corrections (at the level of squared amplitudes) include the interference between FF(2)+2V(2) and LO diagrams, the squares of FF(1) and 2V(1) and their corresponding interference.
The calculation was performed using the Mathematica packages FeynArts \cite{Hahn:2000kx} and FeynCalc \cite{Mertig:1990an} for the generation of the diagrams and the manipulation of amplitudes, and the algorithm FIRE \cite{Smirnov:2008iw} to reduce the resulting expressions into master integrals, which are obtained from Ref. \cite{Ellis:2007qk}.

The partonic virtual corrections $\sigma_v$ to the cross section are obtained by integrating the squared amplitudes over the Higgs pair phase space, that is
\beq
\sigma_v=\f{1}{2s}\f{1}{2\, 2^2 8^2 (1-\ep)^2}\int 
\left\vert \overline{{\cal M}}\right\vert^2d\text{PS}\,,
\eeq
where we also include the flux factor, the average over helicities 
and colors of the incoming gluons and the factor for identical particles in the final state.
Expanding in powers of the strong coupling $\as$:
\beq
\sigma_v=
\left(\f{\as}{2\pi}\right)^2
\left[
\sigma^{(0)}+\f{\as}{2\pi}\sigma^{(1)}+\left(\f{\as}{2\pi}\right)^2\sigma^{(2)}
+{\cal O}(\as^3)\right]\,.
\eeq
The renormalized NLO virtual contribution $\sigma^{(1)}$ is given by
\beeq\label{sigma1}
\sigma^{(1)}\!\!\!&=&\!\!\!
\int_{t_-}^{t_+} dt\,
\left\{
2\,\text{Re}\left[\bom{I}_g^{(1)}\right]
\frac{d\sigma}{dt}^{(0)}
+\f{d\sigma^{(1)}_{\text{fin}}}{dt}
\right\}
\,,
\eeeq
while the renormalized NNLO virtual term $\sigma^{(2)}$ can be expressed in the following general way:
\beeq\label{sigma2}
\sigma^{(2)}\!\!\!&=&\!\!\!
\int_{t_-}^{t_+} dt\,
\left\{
\left(
\left|\bom{I}_g^{(1)}\right|^2 
+2\text{Re}\left[\left(\bom{I}_g^{(1)}\right)^2\right]
+2\text{Re}\left[\bom{I}_g^{(2)}\right]\right)
\frac{d\sigma}{dt}^{(0)}\!\!
+
2\,\text{Re}\left[\bom{I}_g^{(1)}\right]\f{d\sigma^{(1)}_{\text{fin}}}{dt}
+\f{d\sigma^{(2)}_{\text{fin}}}{dt}\,\right\},
\eeeq
where we have used Catani's formula for the infrared singular behaviour of the two-loop QCD amplitudes \cite{Catani:1998bh,Sterman:2002qn,Aybat:2006wq},
and we have defined the quantities:
\beeq
\f{d\sigma}{dt}^{(0)}\!\!\!\!=
F_{\text{LO}}
|C_{\text{LO}}|^2 (1-\ep),
\;\;\;\;\,\,&&\;\;
C_{LO} = \f{6\, \lambda\, v^2}{s-M_H^2+ i M_H \Gamma_H}-1,\nn\\
F_{\text{LO}}=\f{G_F^2}{2304\pi(1-\ep)^2}\,f(\ep)
\,,
\;\;&&\;\;
f(\ep)=\f{1}{\Gamma(1-\ep)}\left[
\f{s(s-4M_H^2)-(t-u)^2}{16\pi s}
\right]^{-\ep}
\,.
\eeeq
All the dependence on the Higgs trilinear coupling $\lambda$ is embodied in the coefficient $C_{\text{LO}}$.
The explicit expression for the one-loop $\bom{I}_g^{(1)}$ and two-loop $\bom{I}_g^{(2)}$ insertion operators  can be found in Ref. \cite{Catani:1998bh}. We recall here that they are functions of the dimensional regularization parameter $\ep$, with poles up to $1/\ep^2$ and $1/\ep^4$, respectively.
The function $f(\ep)$ originates in the $n$-dimensional two-particle phase space, and verifies that $f(0)=1$.

While the singular behaviour of the two-loop amplitudes can be anticipated, the finite contributions $\sigma_{\text{fin}}$ can only be obtained after performing the full two-loop calculation. 
The pole structure of our result agrees with the expressions in Eqs.(\ref{sigma1}) and (\ref{sigma2}) and the infrared-finite contributions for Higgs pair production can be cast into the form
\beeq
\f{d\sigma^{(1)}_{\text{fin}}}{dt}&=&F_{\text{LO}}
\left\{
|C_{\text{LO}}|^2\,{\cal F}^{(1)}
+\text{Re}(C_{\text{LO}})\,{\cal R}^{(1)}
+{\cal O}(\ep^3)
\right\}\,,\\
\f{d\sigma^{(2)}_{\text{fin}}}{dt}&=&F_{\text{LO}}
\left\{
|C_{\text{LO}}|^2 \,{\cal F}^{(2)}
+\text{Re}(C_{\text{LO}}) 
\,{\cal R}^{(2)}
+\text{Im}(C_{\text{LO}}) \,{\cal I}^{(2)}
+ \,{\cal V}^{(2)}
+{\cal O}(\ep)
\right\}\,.\nn
\eeeq
For simplicity, we set $\mu_R^2=s$ in the following expressions. We find that the one-loop contributions are given by
\beeq
   {\cal R}^{(1)}&=&
\frac{4}{3}-\ep
\left[\f{4 M_H^2}{3s}  -\f{2 M_H^4}{3s} \left(\f{1}{t}+\f{1}{u}\right) + \f{2}{3}\right]\,,\\
{\cal F}^{(1)}&=&
11
+\ep
   \left(\frac{7}{6} \zeta_2
   (2 N_f-33)+12 \zeta_3-17\right)\nn\\
   &+&\ep^2 \bigg(
\frac{7}{6} \zeta_2 (33-2 N_f)+   
   \frac{1}{9} \zeta_3
   (2 N_f-141)
   +18\zeta_4-12\bigg)\,.\nn
\eeeq
The expansion of $\sigma^{(1)}_{\text{fin}}$ is needed up to order $\ep^2$ because of the double poles present in $\bom{I}_g^{(1)}$.
The ${\cal F}^{(1)}$ contribution arises from the interference between FF(1) and LO, while  ${\cal R}^{(1)}$ originates from the interference of 2V(1) with the LO. The expansion up to ${\cal O}(\ep^0)$ agrees with the result presented in Ref.\cite{Dawson:1998py}.

The two-loop infrared regulated contributions take the following form:
\beeq
{\cal V}^{(2)}&=&
\f{1}{(3 s t u)^2}\left[
M_H^8 (t+u)^2-2 M_H^4 t u (t+u)^2+t^2 u^2 \left(4 s^2+(t+u)^2\right)
\right]\,,
\\
{\cal I}^{(2)}&=&
4\pi  
\left(1+\f{2 M_H^4}{s^2}\right)
\log\left(
\f{(M_H^2-t)(M_H^2-u)}{t\, u}
\right)
\,,
\\
{\cal F}^{(2)}&=&
\left(\frac{8 N_f}{3}+\frac{19}{2}\right) \log\left(\f{s}{M_t^2}\right)
+N_f \left(\frac{217
   \zeta_2}{12}-\frac{17
   \zeta_3}{6}-\frac{3239}{108}\right)\\
&-&\frac{11  \zeta_2 N_f^2 }{18}
   -\frac{249
   \zeta_2}{2}-\frac{253 \zeta_3}{4}+\frac{45
   \zeta_4}{8}+\frac{8971}{36}\,,\nn
\eeeq
\beeq
{\cal R}^{(2)}&=&-
 \left(1+\f{2 M_H^4}{s^2}\right)
\left\{
-\f{24}{3}\zeta_2
+2\text{Li}_2\left(1-\f{M_H^4}{t\,u}\right)
+4\text{Li}_2\left(\f{M_H^2}{t}\right)
+4\text{Li}_2\left(\f{M_H^2}{u}\right)\right.
\\
&+&\left.
4\log\left(1-\f{M_H^2}{t}\right)\log\left(-\f{M_H^2}{t}\right)
+4\log\left(1-\f{M_H^2}{u}\right)\log\left(-\f{M_H^2}{u}\right)
-\log^2\left(\f{t}{u}\right)
\right\}\nn\\
   &+&\f{4 M_H^2}{s}+\f{314}{9}-\f{20}{27} N_f
- \f{33-2 N_f}{9}  \log
   \left(\frac{t\, u}{s^2}\right)
+8 (C_{H}^{(2)}-C_{HH}^{(2)})      
   \nn\,.
\eeeq
Here ${\cal F}^{(2)}$ originates  from the interference between the two-loop form factor-like diagrams FF(2) and LO contribution plus the square of FF(1), while ${\cal V}^{(2)}$ arises from the square of the tree-level diagram 2V(1). The terms ${\cal R}^{(2)}$ and ${\cal I}^{(2)}$ combine the contributions of two interferences: 2V(2) with LO, and 2V(1) with FF(1).
The Mandelstam variables are given by the expressions:
\beeq
s &=& Q^2\,, \nn\\
t &=& -\f{1}{2}\left[Q^2-2M_H^2-\sqrt{Q^2(Q^2-4M_H^2)}\cos\theta\right]\,,\\
u &=& -\f{1}{2}\left[Q^2-2M_H^2+\sqrt{Q^2(Q^2-4M_H^2)}\cos\theta\right]\,,\nn
\eeeq
while the integration limits $t_\pm$ correspond to $\cos\theta=\pm 1$ and $Q$ is the double-Higgs invariant mass.
The last term in ${\cal R}^{(2)}$, originated on form factor-like contributions, vanishes if the two-loop corrections to the effective vertex $ggHH$ are the same as those of $gHH$.

\section{NNLO Soft-Virtual approximation}

\begin{figure}
\vspace{-0.5cm}
\begin{center}
\begin{tabular}{c}
\epsfxsize=10truecm
\epsffile{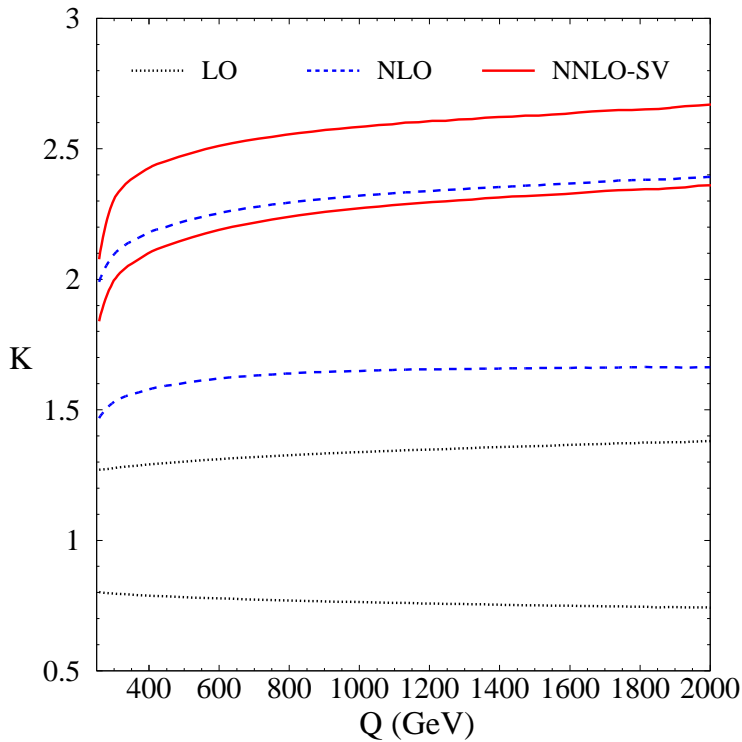}\\
\end{tabular}
\end{center}
\vspace{-0.8cm}
\caption{\label{Kfactors}
{\it K}-factors for Higgs pair production at the LHC as a function of the Higgs pair invariant mass $Q$. 
The bands are obtained by varying the renormalization and factorization scales as described in the main text.}
\end{figure}

Expressed as in Eq.(\ref{sigma2}), the (finite parts of the) two-loop corrections are ready to be implemented in the NNLO soft-virtual (SV) approximation universal formula derived in Ref. \cite{deFlorian:2012za}.
We do not attempt for a full phenomenological analysis of the process at this level, and mostly use the SV approximation as a way to evaluate the impact of the new two-loop results in the cross section.
Therefore, in Figure \ref{Kfactors} we show the NLO and NNLO-SV {\it K}-factors for
 proton-proton collisions at the LHC with c.m. energy $\sqrt{s_H}=14\,\text{TeV}$, in terms of the invariant mass of the Higgs pair.
 Here the NNLO-SV approximation is defined by adding the pure NNLO-SV contribution to the full NLO result.
 At each order, we use the corresponding MSTW2008 set of parton distributions and QCD coupling \cite{Martin:2009iq}.
The bands are obtained by independently varying the scales $\mu_R$ and $\mu_F$ in the range $0.5\, Q\leq \mu_R,\mu_F \leq 2\, Q$, with the constraint $0.5\leq \mu_R/\mu_F \leq 2$. The LO cross section that normalizes the {\it K}-factors is computed at $\mu_R=\mu_F=Q$.
We recall that we always rely on the heavy top quark limit,
and that we use the SV approximation as defined in Mellin space.
As mentioned before, since the coefficient $C_{HH}^{(2)}$ is still unknown we assume $C_{HH}^{(2)}=C_{H}^{(2)}$ for the numerical results.

As can be seen from the plot, we find a large {\it K}-factor, with $K_{\text{NNLO}}^{\,\text{SV}}=2.37$ for the total cross section, resulting in an increase of $23\%$ with respect to the previous order ($K_{\text{NLO}}=1.92$). This value remains approximately constant along the entire Higgs pair invariant mass distribution, with the exception of the region near the threshold where the cross section is anyway very small. Despite of the still sizable corrections, it is noticeable the improvement in the perturbative expansion in the strong coupling constant, which shows the first signs of convergence at NNLO. It is only at this order than there is a (yet not very significant) overlap between two consecutive scale dependent bands.
We can also observe that the scale dependence is substantially reduced: the NNLO band results in a about a $\pm 8\%$ variation around the central value, more than a factor of two smaller than the corresponding NLO band.

We want to recall that in the case of single-Higgs boson production the soft-virtual approximation (compared to the full NNLO result) is known to be accurate to a few percent level. We expect it to be even better for Higgs pair production due to the larger invariant mass of the final state, which leaves less energy for extra hard radiation.
In fact, we computed the NLO soft-virtual cross section, finding $K_{\text{NLO}}^{\,\text{SV}}=1.95$, which differs from the full NLO result by less than $2\%$.
In contrast, the heavy top quark approximation is not expected to be as good as for single-Higgs production since the invariant mass of the Higgs pair is not small compared to the top quark mass. Still a number of improvements can be applied to the current approximation, like keeping the exact full mass dependent LO expressions wherever they appear in the higher order expansion \cite{Dawson:1998py}.
Future work may be directed either towards a full NNLO calculation (in the heavy top limit), or to compute subleading terms in the heavy top quark mass expansion.

\section*{Acknowledgements}

We would like to thank M. Spira for useful discussions.
This work was supported in part by UBACYT, CONICET, ANPCyT and the Research Executive Agency (REA) of the European Union under the Grant Agreement number PITN-GA-2010-264564 (LHCPhenoNet).

\end{document}